# Clot Treatment via Compression- and Shear-Induced Densification of Fibrin Network Microstructure: A Combined in Vitro and In Silico Investigation

**Authors and Affiliations**

Yilong Chang[a, 1], Guansheng Li[b, 1], Jay Sim[a], George Em Karniadakis[b, *], Ruike Renee Zhao[a, *]

[a]Department of Mechanical Engineering, Stanford University; Stanford, CA 94305, USA.

[b]Division of Applied Mathematics, Brown University, Providence, RI 02912, USA.

[1]These authors contribute equally: Yilong Chang, Guansheng Li

**\*Corresponding authors**

George Em Karniadakis (george_karniadakis@brown.edu)

Ruike Renee Zhao (rrzhao@stanford.edu)

**Abstract**

Blood clots, consisting of red blood cells (RBCs) entrapped within a fibrin network, can cause life-threatening conditions such as stroke and heart attack. The recently developed milli-spinner thrombectomy device presents a promising mechanical approach to removing clots by substantially modifying the microstructure of the blood clot, resulting in up to 95% volume reduction through combined compressive and shear forces. To better understand the mechanism and optimize this approach, it is important to quantitatively understand of how compression and shear loadings alter the clot structure. In this study, we combine in vitro experiments with dissipative particle dynamics (DPD) simulations to investigate the effectiveness of clot debulking under integrated compression and shear. Controlled experiments quantify clot volume changes, while simulations offer microscopic insight into fibrin network densification and RBC release. This integrated approach enables a systematic evaluation of mechanical response and microstructure change of different clot types, providing fundamental knowledge to guide the rational design of next-generation mechanical thrombectomy technologies.

## Introduction

Blood clots can form in various locations throughout the body, leading to vessel occlusion and potentially life-threatening conditions such as strokes and heart attacks (1-3). A clot is primarily composed of red blood cells (RBCs) trapped within a fibrin network (**Fig. 1A(i)**) (4, 5). The milli-spinner thrombectomy, a recently developed technology, removes clots through a novel mechanism that densifies the clot microstructure to facilitate efficient extraction (6-8). It has demonstrated high effectiveness by significantly reducing clot volume and leaving a compact and highly densified fibrin core that can be easily removed (**Fig. 1A(ii)**). During operation, as illustrated in **Fig. 1B**, the milli-spinner's unique structural design induces coupled compression and shear forces on the clot, which compacts the fibrin network and releases RBCs. This mechanism leads to substantial volume reduction up to 95% and enables rapid clot removal, often in under a minute. The visible transition from a red clot to a white dense fibrin core is observed, which can be easily aspirated into the catheter. Scanning electron microscope (SEM) images (**Fig. 1C**) further validate the densification of the clot microstructure, showing a transformation from the pristine clot with a sparse fibrin network and rich with RBCs, to a post-treatment structure composed of a dense fibrin network without RBCs. The milli-spinner thrombectomy represents a promising direction for more effective clot treatment. To support and further optimize its development, a quantitative understanding of how compression and shear forces synergistically transform the clot microstructure to drive volume reduction is essential.

In this study, we employ a combined in vitro and in silico approach to investigate how integrated compression and shear forces affect clot debulking efficiency. In the in vitro experiments, clots are subjected to independently controlled compression and shear loads, and resulting volume changes are measured to evaluate the debulking efficiency. In parallel, in silico studies using microscopic dissipative particle dynamics (DPD) modeling are conducted to quantitatively analyze and visualize the fibrin network densification and RBC release process during the compression- and shear-induced clot debulking (9-11). Through this combined methodology, we systematically evaluate how varying compression and shear affect clot volume reduction across different clot compositions, including fibrin clots and clots with varying RBC content.

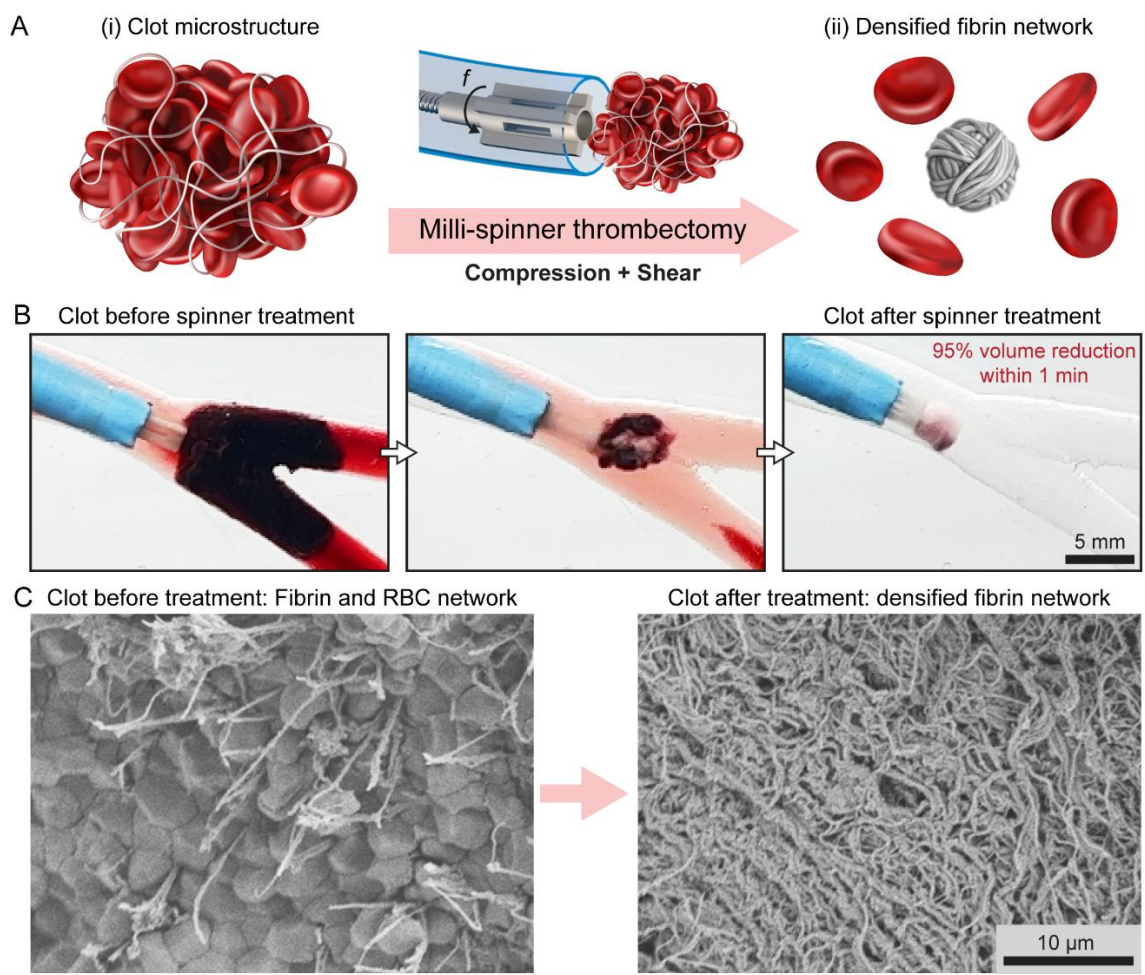

**Fig. 1. Milli-spinner debulks a clot through compression and shear.** (**A**) Schematic illustration of the debulking mechanism, showing how integrated compression and shear forces densify the fibrin network and facilitate RBC release. (**B**) Experimental demonstration of the milli-spinner debulking a clot, showing significant volume reduction and a visible color change from red (RBC-rich) to white (fibrin-dense). (**C**) SEM images of clot microstructure before and after milli-spinner treatment, highlighting the transition from a loose fibrin network with embedded RBCs to a compact fibrin structure.

## Results

### Combined in vitro and in silico approaches for clot-debulking investigation

To systematically investigate how combined compression and shear influence clot debulking, we design an experimental setup that allows independent control of the compressive and shear forces applied to the clots. As illustrated in **Fig. 2A**, the setup consists of a rotating disk in direct contact with the clot, which is confined within a transparent tube. This configuration enables the simultaneous application of controlled compression and shear (see **SI Appendix, Fig. S1** and **section S1** for experimental setup and procedure details). The clot is molded into a cylindrical shape, with a diameter closely matching the diameter of the testing tube (14.6 mm) and a height of approximately 12 mm (see **Materials and Methods** for clot fabrication). The rotating disk is mounted to a universal testing machine, which provides precise control of compression $p$ applied to the clot through vertical displacement. Shear is introduced by rotating the disk against the clot, with the shear input characterized by a defined rotational frequency $f$. The clot remains stationary due to frictional resistance, allowing shear to be transmitted across the interface for debulking. However, the actual shear force transmitted to the clot is difficult to quantify directly, as it depends on factors such as the clot surface properties and the interfacial characteristics between the clot and the disk. The disk applies compression to the clot at a defined pressure and simultaneously rotates concentrically to apply shear. The resulting clot volume reduction is quantified over time as the fibrin network densifies. The percentage of volume reduction, denoted as $\Delta V$, is calculated as $(L_0 - L)/L_0$, where $L_0$ is the initial clot length, and $L$ is the current clot length (see **SI Appendix, section S2** for calculation details). **Fig. 2B** illustrates a representative example. A clot with 5% RBC content is subjected to 8 kPa compression while the disk rotates at 4k rpm. After 25 s, the clot shows a 30.0% volume reduction, and eventually reaches a 79.6% volume reduction after 250 s.

To quantitatively study the microscopic fibrin network densification induced by compression and shear, which leads to the dramatic clot volume reduction, a DPD model is established (**Fig. 2C**) (9-12). The simulation system comprises a disk, a cylindrical tube, and a clot composed of a fibrin network embedded with RBCs. The clot is confined within a cylindrical domain (20 μm in diameter, 40 μm in height), while a 10 μm-diameter disk applies compression and shear simultaneously to debulk the clot. The DPD framework incorporates conservative, dissipative, and random forces to model interactions among fluid, RBCs, fibrin fibers, and the tube wall (see **SI Appendix, section S3** for DPD and RBC models) (13-15). A calibrated phenomenological bilinear force–strain model is employed to characterize the mechanical behavior of fibrin fibers during the densification process (see **SI Appendix, section S4** for details) (16-18). The disk-clot interfacial property used in simulations is calibrated through matching the simulated volume reduction with experimental measurements, which is detailed in *Quantitative evaluation of fibrin clot debulking by compression and shear* section. As illustrated in **Fig. 2D,** when a 5% RBC clot is interacting with the spinning disk at 4k rpm with a compression of 8 kPa, fibrin network densification and RBC release are clearly observed during the clot debulking process, eventually resulting in a volume reduction of 77.0% at simulation (see **SI Appendix, section S2** for volume reduction calculations). These simulation results show good agreement with experimental measurements, validating the DPD model's capability to link clot microstructure changes with its macroscopic debulking behavior.

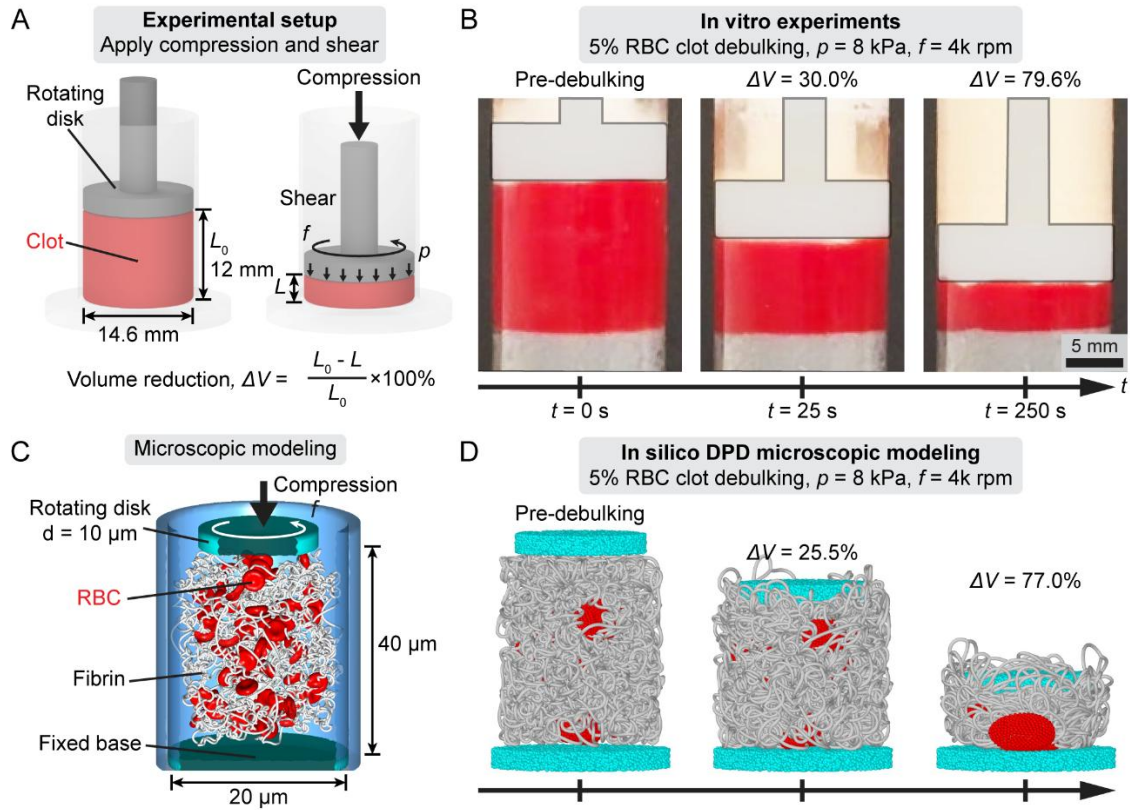

**Fig. 2. Combined in vitro and in silico approaches for clot debulking investigation. (A)** Schematic of the experimental setup applying compression and shear, where a clot is confined in a cylindrical tube and subjected to controlled compression and shear by a rotating disk. (**B**) Experimental demonstration of debulking a 5% RBC clot under 8 kPa compression and disk rotational frequency of 4k rpm. The volume reduction $\Delta V$ reaches 30.0% at $t = 25$ s and 79.6% at $t = 250$ s. (**C**) DPD model of a microscale clot confined

within a cylindrical tube. The clot is subjected to the same loading condition as the experiment. (**D**) Simulated debulking of a 5% RBC clot, exhibiting fibrin densification, RBC release, a volume reduction of 25.5% and a final volume reduction of 77.0%.

**Quantitative evaluation of fibrin clot debulking by compression and shear**

We begin by using fibrin clots (without RBCs) to systematically and quantitatively investigate the effect of compression and shear on clot debulking (see **Methods** for fibrin clot fabrication). **Fig. 3A** shows a comparison of fibrin clots before and after debulking (**Movie S1**). A cylindrical fibrin clot is subjected to a compression of 8 kPa and sheared by a disk rotating at 4k rpm. This treatment results in a substantial clot volume reduction, reaching 80.0% after 178 s. The SEM images of the fibrin clot before and after debulking demonstrate the underlying mechanism of the clot volume reduction, which is the densification of the fibrin network (**Materials and Methods**). The initial porous fibrin network becomes significantly compacted under combined compression and shear loading. Simulation of the fibrin densification process captures the microstructure transformation observed in the experiment (**Fig. 3B** and **Movie S1**). The modeled fibrin clot is constructed based on experimental measurement to match the fibrin volume fraction in the fibrin clot (see **SI Appendix, section S5** for details). Under the same loading conditions ($p = 8$ kPa, $f = 4$k rpm), the simulation demonstrates an 81.3% volume reduction, consistent with the experimentally observed fibrin network densification. Both experimental and simulation results confirm that fibrin network densification is the primary mechanism driving fibrin clot volume reduction under combined compression and shear loading.

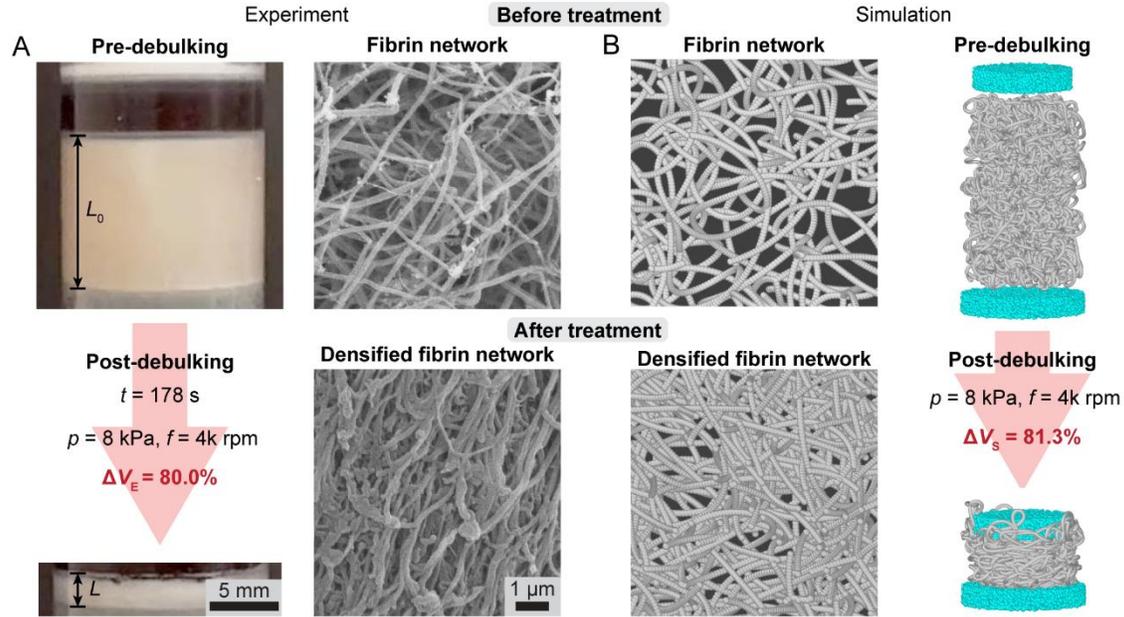

**Fig. 3 Experimental and simulation results on fibrin clot debulking.** (**A**) Experimental comparison of fibrin clot pre-debulking and post-debulking, showing a volume reduction $\Delta V$ of 80.0% after 178 s of loading ($p$ = 8 kPa, $f$ = 4k rpm). The SEM images of clot microstructure pre-debulking and post-debulking demonstrate fibrin densification. (**B**) Simulated comparison of fibrin clot pre-debulking and post-debulking under the same load conditions ($p$ = 8 kPa, $f$ = 4k rpm), showing evident fibrin densification and a volume reduction $\Delta V$ of 81.3%. The initial porous fibrin network is densified into a dense fibrin core.

To investigate the individual contributions of compression and shear to fibrin clot debulking, we decouple the loading modes. First, we fix the disk rotational frequency at $f$ = 4k rpm (**Fig. 4A**) and systematically vary the applied compression at $p$ = 4, 8, 12, and 16 kPa. Experimentally, as the applied pressure increases, the time required to achieve a 50% volume reduction decreases from 85 s to 25 s, indicating that higher pressure facilitates faster fibrin clot densification. Additionally, the final volume reduction increases with pressure, rising from 73.8% at 4 kPa to 81.6% at 16 kPa. Under the same conditions, simulations are conducted showing a similar final volume reduction to the experiment. At 4 kPa, the simulated volume reduction reaches 77.0% and increases to 88.3% at 16 kPa. Both experimental and simulated results show that an increase in compression pressure enhances both the rate and magnitude of fibrin clot volume reduction. Notably, the improvements are non-linear and begin to plateau at higher pressures (8 kPa), suggesting that effective clot debulking can be achieved without the need for excessively high compressive loads.

The effect of shear on fibrin clot debulking is then investigated by maintaining compression while varying the disk rotational frequency to generate different levels of shear (**Fig. 4B**). Under pure compression ($f = 0$ rpm), a fibrin clot subjected to 8 kPa compression reaches a 55.2% volume reduction after 400 s in the experiment, while the simulated clot reaches 63.3%. Impressively, introducing only a small shear through a low disk rotational frequency ($f = 500$ rpm) significantly enhances both the rate and magnitude of the volume reduction, achieving 75.1% and 72.9% reduction in experiment and simulation, respectively. This improvement highlights the critical role of shear in facilitating fibrin network densification. As the rotational frequency increases further to 4k and 6k rpm, the final volume reduction continues to rise but begins to plateau, reaching between 79.3% to 86.4% across experiments and simulations. Comparing the results from decoupled compression and shear tests, the combination of compression and shear is clearly demonstrated to be essential for achieving rapid and substantial fibrin clot volume reduction.

The properties of clinical fibrin clots, particularly fibrin content, depend on the clot's age and location of clot formation (19-21), which influence the mechanical response of clots to compression and shear forces (22). Thus, it is essential to investigate how clots with different fibrin content respond under coupled compression and shear loads. Experimentally, it is challenging to form clots with precisely controlled fibrin content in vitro as it depends on the fibrinogen level in blood (23, 24). Here, simulations are used to systematically investigate the debulking behavior of fibrin clots with varying fibrin content (1%, 2%, 3%, and 4% volume fraction) under compression and shear ($p = 8$ kPa, $f = 4$k rpm) (see **SI Appendix, section S5** for fibrin clot model details). The results show that clots with higher fibrin content exhibit both slower volume reduction and lower final volume reduction compared to clots with lower fibrin content (**Fig. 4C**). As the fibrin content increases from 1% to 4%, the final volume reduction decreases from 93.3% to 81.3%. Experimental validation using a measured 4% fibrin clot achieves an 80.0% volume reduction under the same loading condition, which is consistent with the simulation results. These findings indicate that, under the same compression and shear loading conditions, clots with a higher fibrin content have a lower debulking efficiency.

The shear force applied on clots arises from the friction between the rotating disk and the clot, which is influenced by the surface property of the disk, and, consequently, affects the clot debulking behavior. To evaluate this effect, we perform DPD simulations to quantitatively assess the impact of disk–clot friction on the debulking process (12, 25, 26). In the DPD framework, the total force $f_i$ exerted on particle i by particle j is composed of a conservative force $F_{ij}^C$, the dissipative force $F_{ij}^D$, and a random force $F_{ij}^R$. The dissipative force is directly related to the friction between disk and fibrin, given by $F_{ij}^D = \gamma \omega_d(r_{ij})(\vec{r_{ij}} \cdot \vec{v_{ij}})\vec{r_{ij}}$, for $r_{ij} \leq r_c$, where $r_c$ is a cut-off radius, $\gamma$ is the dissipative coefficient $r_{ij}$ is the distance between two particles with the corresponding unit vector $\vec{r_{ij}}$, $and\ \vec{v_{ij}}$ is the relative velocity. To capture the effects of varying disk surface properties on clot debulking, the dissipative coefficient $\gamma$ is systematically varied from 0 to 250 in simulations performed under constant compression (8 kPa) and rotational frequency (4k rpm), representing increasing disk-clot friction. The value of $\gamma$ used in the simulation is calibrated by fitting it to match the experimental measurements of clot debulking volume reduction. As shown in **Fig. 4D**, increasing $\gamma$ enhances the volume reduction rate and clot volume reduction. When $\gamma = 250$, the simulated final volume reduction of 81.3% matches with the experimentally measured results of 80.0% (star symbol in **Fig. 4D**). Therefore, $\gamma = 250$ is adopted for all the simulations in this study.

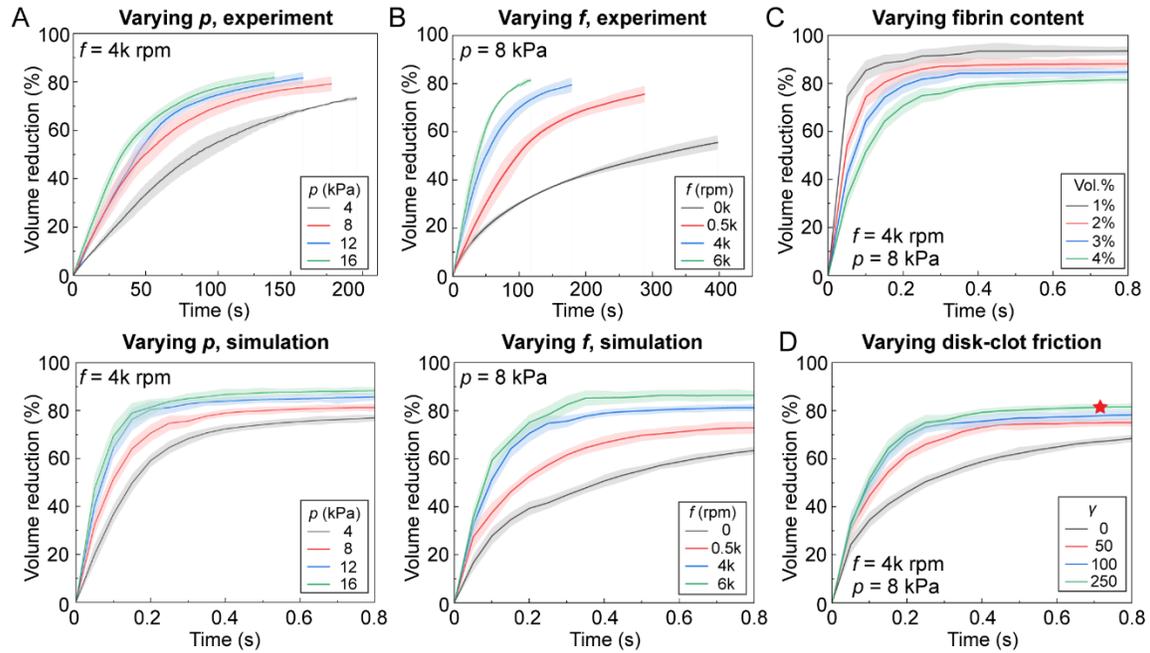

**Fig. 4 Systematic study of fibrin clot debulking under varying compression, shear, clot fibrin, and disk-clot friction.** Fibrin clot volume reduction for (**A**) varying compression ($p$ = 4, 8, 12, and 16 kPa) at a fixed disk rotational frequency of 4k rpm in the experiment. The shaded regions represent two standard deviations calculated from 3 repeated tests. (**B**) Varying disk rotational frequency ($f$ = 0, 500, 4k, and 6k rpm) at a fixed compression of 8 kPa in the experiment. (**C**) Varying fibrin content (1%, 2%, 3%, and 4% in volume) under constant loading ($p$ = 8 kPa, $f$ = 4k rpm) in simulation. (**D**) Disk-clot friction characterized by different dissipative coefficients ($\gamma$ = 0, 50, 100, and 250), under constant compressive and shear loading ($p$ = 8 kPa, $f$ = 4,000 rpm) in simulation.

**Quantitative evaluation of RBC clot debulking induced by compression and shear**

Clinical clots exhibit considerable compositional variability, particularly in their RBC content, which can range from 0% (fibrin clot) to over 80% (RBC-rich clot) (27-29). This variability leads to significantly different mechanical responses under the same applied loading conditions, which in turn critically influences the effectiveness of endovascular thrombectomy procedures. To investigate how RBC content influences clot debulking behavior during milli-spinner thrombectomy, we perform a systematic investigation to assess the effects of combined compression and shear on debulking clots with varying RBC contents.

To accurately quantify fibrin content and construct RBC clot models for simulation, a two-dimensional (2D) projection method is employed to assess fibrin content in clots containing 5% and 30% RBCs (see **SI Appendix, section S6** for details on modeling of RBC clots in simulation). The constructed RBC clot models demonstrate strong visual agreement with the experimental SEM images (**Fig. 5**). **Fig. 5A** and **B** show the microstructural changes of the surface of clots (5% and 30% RBCs, respectively) before and after debulking under compression and shear ($p$ = 8 kPa, $f$ = 4k rpm) in the experiment and simulation. SEM images of both 5% and 30% RBC clots after debulking reveal fibrin network densification, consistent with observations from simulations (see **SI Appendix, Fig. S8** for the clot inner microstructure). Notably, unlike fibrin clots, where volume reduction arises solely from fibrin network densification, RBC clots exhibit volume reduction through a combination of fibrin densification and RBC release from the clot fibrin network. In clots with 5% RBC content, simulated clot debulking is predominantly driven by fibrin densification, with some RBCs remaining entrapped within the densified clot network. In contrast, clots with 30% RBC content undergo both fibrin densification and RBC release from the clot network. (see **Movie S2**).

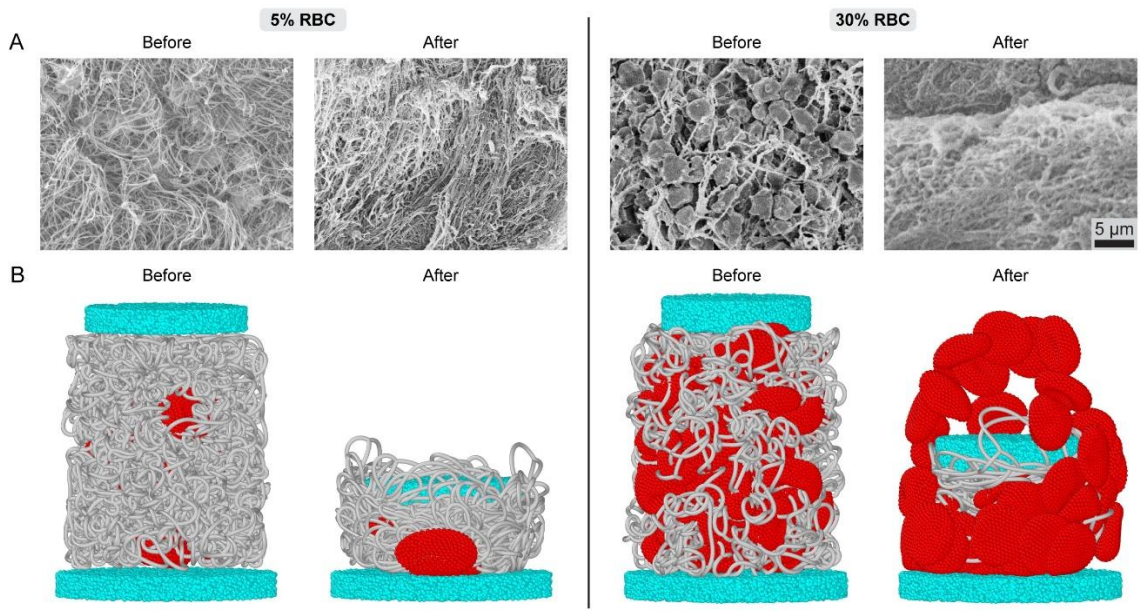

**Fig. 5 Experimental and simulation results on debulking of clots with varying RBC contents.** (**A**) SEM images of clots with 5% and 30% RBCs, highlighting microstructural changes before and after debulking under compression and shear ($p$ = 8 kPa, $f$ = 4k rpm).

(**B**) 3D simulation results depicting fibrin network densification and RBCs release in clots with 5% and 30% RBCs under compression and shear loading ($p$ = 8 kPa, $f$ = 4k rpm).

Volume reduction of clots with 5%, 10%, and 30% RBC content subjected to compression and shear ($p$ = 8 kPa, $f$ = 4k rpm) are characterized and shown in **Fig. 6A**. Both experimental and simulation results show a clear trend suggesting that as the RBC content in the clot increases, the final clot volume reduction decreases. Specifically, as the RBC content in the clot increases from 5% to 30%, the final volume reduction decreases from 78.1% to 52.7% in the experiment, corresponding to 77.0% and 60.2% volume reduction in simulation. The debulking of RBC clot exhibits a two-stage process in simulation: the first stage involves rapid fibrin network densification, leading to a fast initial volume reduction, while the second stage corresponds to the gradual release of RBCs from the fibrin network, resulting in a slower volume reduction rate. As previously discussed, the volume reduction of clots with 5% RBC is primarily driven by fibrin networking densification, whereas for the 30% RBC clot, the volume reduction results from a combination of fibrin densification and RBC release. In the case of a 30% RBC clot, although some RBCs are released from the fibrin network and contribute partially to volume reduction, this process is more difficult to achieve compared to fibrin network densification. Simulation results support this observation, showing that some RBCs remain trapped within the fibrin network following debulking.

It is important to note that the actual milli-spinner thrombectomy system can achieve up to 95% clot volume reduction in a short time, due to differences in operating conditions compared to the current experimental setup. During milli-spinner operation, clots with high RBC content debulk more rapidly and extensively, driven by clot reorientation that enables shear to act across all surfaces of the clot. In contrast, the current setup applies shear to only one surface, limiting RBC release and impeding fibrin network densification. As a result, clots with higher RBC content exhibit lower volume reduction in this setup

To examine how compression and shear individually influence RBC clot volume reduction, we first fix the disk rotational frequency at 4k rpm and systematically increase the applied compression pressure from 1 kPa to 16 kPa (**Fig. 6B**). At low compression of 1 kPa, the 10% RBC clot exhibits slower volume reduction, reaching 35% in 250 s. As the applied compression increases, both the rate and magnitude of volume reduction improve, achieving 80.9% at 16 kPa in the experiment and 79.2% in the simulation. Both experimental and simulation results demonstrate a non-linear relationship between the increase in compression and volume reduction. Compared to fibrin clots under the same loading conditions, the 10% RBC clot exhibits faster initial volume reduction due to densification of the relative spare fibrin network but ultimately achieves a lower final volume reduction, indicating that RBC release is more difficult to achieve than fibrin densification.

To evaluate the role of shear in clot volume reduction, we fix the compression at 8 kPa and gradually increase the rotational frequency from 0 to 6k rpm (**Fig. 6C**). Under pure compression ($f = 0$k rpm), both experiment and simulation show minimal volume reduction, a phenomenon that contrasts sharply with the behavior observed in fibrin clot under the same loading condition. However, when a smaller amount of shear ($f = 2$k rpm) is introduced, the clot exhibits a marked increase in volume reduction reaching nearly 70% in both experiment and simulation. This demonstrates that RBC release is primarily driven by shear. As the rotational frequency increases further from 2k rpm to 6k rpm, both the rate and magnitude of volume reduction continue to rise but eventually plateau at a higher frequency. While the final clot volume reduction increases slightly with rotational frequency, the changes become less significant at higher speeds. Compared to fibrin clots under the same loading conditions, the final volume reduction of 10% RBC clot remains lower, as a portion of RBCs remains trapped within the fibrin network.

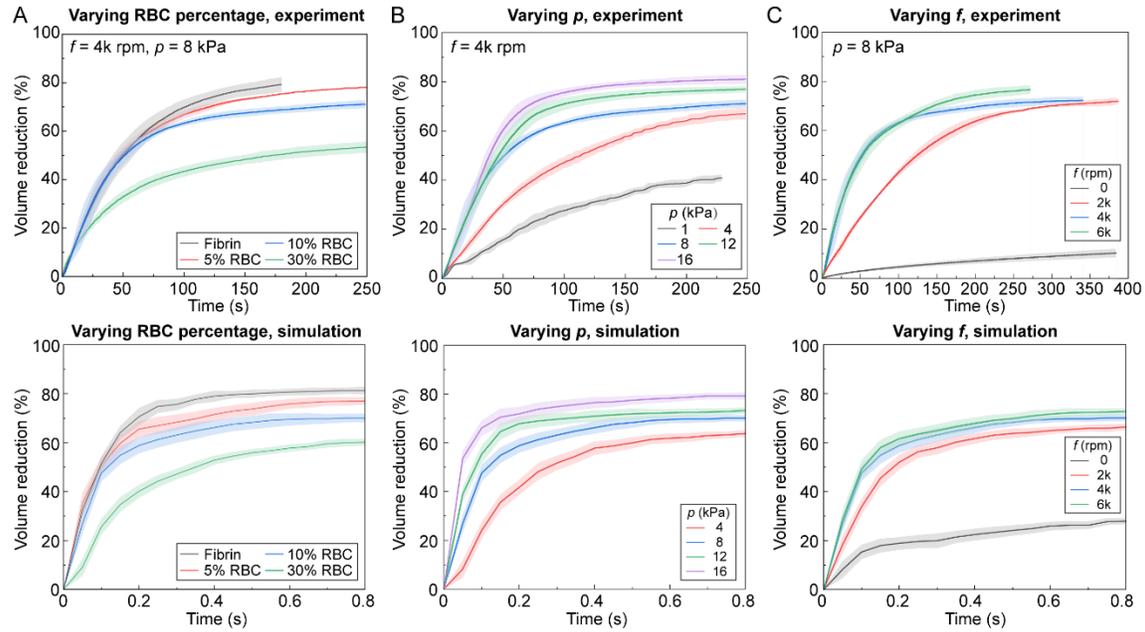

**Fig. 6 Systematic and quantitative study of debulking of clots with varying RBC content under varied compression and shear.** (**A**) Volume reduction of clots with 5%, 10%, and 30% RBC content under constant loading ($p$ = 8 kPa, $f$ = 4k rpm). (**B**) Volume reduction of 10% RBC clot under varied compression pressure ($p$ = 1, 4, 8, 12, and 16 kPa) with a fixed rotational frequency of 4k rpm. (**C**) Volume reduction of 10% RBC clot under varying disk rotational frequency ($f$ = 0, 2k, 4k, and 6k rpm) with a fixed compression of 8 kPa.

## Conclusion

In this study, we investigated compression- and shear-induced clot debulking through a combination of experiments and simulations. By independently controlling the applied compression and shear, we characterized the volume reduction of both fibrin clots and RBC clots. Both the rate and magnitude of volume reduction increase with applied compression and shear. This relationship is non-linear, plateauing at high loads. When a rotating surface is in contact with the clot, compression is necessary to induce shear to create volume reduction. Conversely, under a fixed compression, the application of shear, through the rotating surface, plays a critical role in clot volume reduction, particularly for RBC clot where shear facilitates RBC release. Furthermore, clinical fibrin clot properties, especially fibrin content, vary with clot age and formation site, affecting their mechanical response to compression and shear. We performed simulations evaluating the effect of fibrin content on debulking efficiency and found that clots with higher fibrin content exhibit reduced volume reduction. We also examined the role of disk–clot friction, which governs shear force generation at the clot interface. DPD simulations reveal that increased friction between the clot and disk enhances both the rate and extent of volume reduction. Unlike fibrin clots, where volume reduction is primarily driven by fibrin network densification, volume reduction in RBC clots arises from a combination of fibrin densification and RBC release. These findings provide insights into how compression and shear transform the clot structure and can provide guidance on the development of next-generation mechanical thrombectomy technologies that leverage clot microstructure modification to enhance clot removal.

Several open issues remain for future investigation to build upon the insights from this study. One key area is clot fragmentation under combined compression and shear. Characterizing the mechanical thresholds that lead to clot fracture will be critical for identifying safe and effective loading conditions during mechanical thrombectomy. Another important direction is the systematic evaluation of clots with more complex and heterogeneous cellular compositions, particularly those containing significant populations of platelets and white blood cells. These components are known to influence clot mechanical properties and structural integrity, and their presence may alter the clot's response to mechanical loading. Understanding how these additional cellular constituents affect the clot debulking dynamics could broaden the applicability of the milli-spinner thrombectomy and help tailor device design and operating strategies for diverse clot phenotypes encountered in clinical settings. Further integration of multiscale simulation approaches that combine continuum-scale models with microscale particle dynamics could also offer predictive insights into clot-device interactions across different clot types and loading regimes. Together, these future studies will advance the mechanistic understanding of clot debulking and inform the development of more effective milli-spinner thrombectomy technologies.

## Materials and Methods

### Clot fabrication

Porcine whole blood anticoagulated with 3.8wt% sodium citrate at a 9:1 volume ratio was obtained from Animal Technologies, Inc. The anticoagulated blood was centrifuged at 1100 × g for 15 minutes in 50 mL centrifuge tubes to separate blood components. The separated RBCs and citrated plasma were extracted individually, and the buffy coat layer was discarded. RBCs and plasma were then recombined at varied volume ratios. The mixtures were poured into cylindrical molds (14.6 mm diameter, 12 mm height) and coagulated by adding 2.06wt% calcium chloride at the 9:1 volume ratio. Clots were incubated in water baths at 37 °C for approximately 1 hour and subsequently stored at 5°C for 12 hours. The following day, the clots were de-molded and transferred into a saline solution for an additional day.

### SEM image procedure for visualizing clot microstructure

The microstructure of clots pre- and post-debulking were visualized using SEM. Clots were fixed in a solution containing 4% Paraformaldehyde and 2% Glutaraldehyde in 0.1 M Sodium Cacodylate buffer for 1 hour at room temperature, followed by storage at 5°C for an additional 24 hours to ensure complete fixation. The fixed clots were then washed three times with 0.1M Sodium Cacodylate buffer and post-fixed in 1% osmium tetroxide (diluted in Sodium Cacodylate buffer) for 1 hour at room temperature. After post-fixation, clots were dehydrated in ethanol series (30%, 50%, 70%, 95%, and twice in 100%). Samples were subsequently critical point dried to preserve the clot structure. A 14 nm thick gold coating was applied to the dried clots using a sputter coater. SEM imaging was performed using a Zeiss Sigma scanning electron microscope (Zeiss Inc., Germany) equipped with a GEMINI electron optical column.

### Data Availability

All study data are included in the article and/or SI Appendix

### Acknowledgments

R.Z. acknowledge the support from the Terman Faculty Fellowship and Gabilan Faculty Fellowship. The authours would like to thank Prof. Xianqiao Wang from the University of Georgia for useful information regarding the fibrin model.